\documentclass[prl, twocolumn, amsfonts, amssymb, amsmath,
               showkeys, nofootinbib, superscriptaddress]{revtex4-2}

\usepackage[english]{babel}
\usepackage[utf8]{inputenc}
\usepackage{mathtools}
\usepackage{physics}
\usepackage{booktabs}
\usepackage{siunitx}
\usepackage{graphicx}
\usepackage{dcolumn}
\usepackage{bm}
\usepackage{xcolor}
\usepackage[pdftex,pdftitle={AGN H0},pdfauthor={Kumar et al.}]{hyperref}
\usepackage{multirow}
\usepackage{hhline}
\usepackage{aas_macros}
\usepackage{orcidlink}
\usepackage[table]{xcolor}
\usepackage{makecell}
\usepackage{subcaption}

\definecolor{scvigreen}{rgb}{0.85, 0.96, 0.85}
\definecolor{revred}{rgb}{0.75, 0.00, 0.00}

\begin{document}
\preprint{APS/123-QED}

\title{Hubble constant measurement with 13 bright standard sirens from binary black hole mergers inside active galactic nuclei}

\author{Dhruv Kumar\orcidlink{0000-0001-8205-0404}}
\affiliation{Beijing Institute of Mathematical Sciences and Applications, Beijing 101408, China}
\affiliation{Department of Physics, National Institute of Technology Agartala, Tripura 799046, India}
\affiliation{Institute for Gravitation and the Cosmos, Department of Physics, Pennsylvania State University, University Park, PA 16802, USA}

\author{Alejandro Torres-Orjuela\orcidlink{0000-0002-5467-3505}}
\email[Corresponding author:]{atorreso@bimsa.cn}
\affiliation{Beijing Institute of Mathematical Sciences and Applications, Beijing 101408, China}

\begin{abstract}
We measure the Hubble constant $H_0$ using 13 gravitational-wave binary black
hole mergers associated with active galactic nucleus hosts. We find
$H_0=70.50^{+3.37}_{-2.89}\,({\rm stat})\pm1.56\,({\rm cal})$\,km\,s$^{-1}$\,Mpc$^{-1}$
($4.4\%$ precision), consistent with both Planck\,2018 ($0.98\sigma$)
and SH0ES\,2024 ($0.76\sigma$), with no significant preference between
the two. Combining with the bright siren GW170817 sharpens the
constraint to $H_0=70.31^{+3.00}_{-2.85}\,({\rm stat})\pm1.55\,({\rm cal})$\,km\,s$^{-1}$\,Mpc$^{-1}$
($4.2\%$ precision), and further combining with an independent dark-and-bright-siren
sample tightens it to $H_0=69.71^{+2.55}_{-2.40}\,({\rm stat})\pm1.54\,({\rm cal})$\,km\,s$^{-1}$\,Mpc$^{-1}$
($3.5\%$ precision). Assuming a luminosity-distance prior centered around the value related to a fixed cosmology in turn, recovers $H_0=67.62\pm0.72$ (Planck-anchored) and
$H_0=72.91\pm0.72$\,km\,s$^{-1}$\,Mpc$^{-1}$ (SH0ES-anchored). We show
that under such an assumption, a rejection of $\gtrsim4\sigma$ to the opposing
anchor is obtained.
\end{abstract}

\maketitle

\textit{Introduction.}---The Hubble constant $H_0$ parametrises the
present-day expansion rate of the Universe, and sits at the centre of a
persistent cosmological tension. The Planck Collaboration's analysis of
the cosmic microwave background (CMB) gives
$H_0=67.4\pm0.5$\,km\,s$^{-1}$\,Mpc$^{-1}$~\cite{Planck2020}, while
the SH0ES team, calibrating Type~Ia supernovae via Large Magellanic Cloud
Cepheids, reports $H_0=73.0\pm1.0$\,km\,s$^{-1}$\,Mpc$^{-1}$~\cite{Riess2022},
a discrepancy exceeding $5\sigma$.
Independent local probes occupy the same contested territory: the tip of
the red giant branch gives
$H_0=69.8\pm0.6\,(\mathrm{stat})\pm1.6\,(\mathrm{sys})$\,km\,s$^{-1}$\,Mpc$^{-1}$~\cite{Freedman2021},
$2\sigma$ below SH0ES; surface brightness fluctuations yield
$H_0=73.3\pm0.7\,(\mathrm{stat})\pm2.4\,(\mathrm{sys})$\,km\,s$^{-1}$\,Mpc$^{-1}$~\cite{Blakeslee_2021},
consistent with SH0ES; and the H0LiCOW strong-lensing analysis finds
$H_0=73.3^{+1.7}_{-1.8}$\,km\,s$^{-1}$\,Mpc$^{-1}$~\cite{H0LiCOW},
$3.1\sigma$ from Planck and $5.3\sigma$ when combined with
SH0ES~\cite{H0LiCOW}. Whether this tension reflects new physics beyond the
$\Lambda$CDM model or unresolved systematics -- in the calibration of the
local distance ladder, the CMB likelihood, or the assumed cosmological
model -- remains open~\cite{Di_Valentino_2021,Verde2019}.

Gravitational wave (GW) standard sirens offer a fully independent route to $H_0$, free from
the iterative calibration of the distance
ladder~\cite{Schutz1986,Holz_2005,Krolak:1987ofj}. The binary inspiral encodes the
absolute luminosity distance $D_L$ directly in the waveform amplitude;
pairing this with a host-galaxy redshift yields $H_0$ without any
intermediate rungs. The binary neutron star (BNS) merger GW170817, jointly observed with a
short $\gamma$-ray burst and kilonova, provided the first such
measurement: $H_0=70^{+12}_{-8}$\,km\,s$^{-1}$\,Mpc$^{-1}$~\cite{Abbott2017Nature},
later refined to
$H_0=68.8^{+5.2}_{-4.9}$\,km\,s$^{-1}$\,Mpc$^{-1}$~\cite{Hotokezaka2019}.
In the absence of a direct electromagnetic (EM) counterpart, statistical dark-siren methods
associate GW events with galaxy catalogues
probabilistically~\cite{Schutz1986,MacLeod_2008,Del_Pozzo_2012,Gray2020},
or exploit features in the compact-binary mass distribution to break the
mass-redshift
degeneracy~\cite{Taylor2012,Farr_2019,Mastrogiovanni_2021}.
The LVK GWTC-5.0 analysis combining 235 GW candidates with GW170817
and the DES Year~6 galaxy catalogue yields
$H_0=71.0^{+9.0}_{-7.1}$\,km\,s$^{-1}$\,Mpc$^{-1}$~\cite{theligoscientificcollaboration2026gwtc50constraintscosmicexpansion},
a $25.7\%$ improvement over
GWTC-4.0~\cite{theligoscientificcollaboration2025gwtc40constraintscosmicexpansion}
but still at ${\approx}11\%$ fractional precision, limited by sparse
catalogue completeness at $z\gtrsim0.3$. Moreover, this measurement is dominated by the bright standard siren GW170817 and its EM counterpart.

A qualitatively different approach becomes possible when a BBH merger
can be uniquely identified through an EM counterpart.
BBH mergers embedded in active galactic nucleus (AGN) accretion
discs are predicted to produce optical flares through disc perturbation
and reprocessing~\cite{McKernan2019ApJL,Graham_2023}.
Such counterparts supply a spectroscopic host redshift, converting $D_L$
directly into $H_0$ with none of the galaxy-population averaging
inherent to dark-siren
methods~\cite{chen_haster_2022,morton_rinaldi_2023}.
In Ref.~\cite{Kumar2026a},
we performed a Bayesian coincidence-classification analysis of
18 gravitational-wave events from the LVK O3--O4b catalogue against
28 candidate AGN counterparts, jointly accounting for GW
sky-localisation, AGN positional coincidence, and SMBH-induced
environmental redshift corrections. Of the 28 candidate pairs, 21
were found to be positive-to-strong-favoured against the vacuum
baseline (log-Bayes factor $>1$). 13 unique one-to-one BBH--AGN
associations were then identified once competing counterparts for
the same event, or the same AGN claimed by multiple events, were
adjudicated by their relative evidence. In this letter, we exploit
these 13 favoured associations to perform
GW-based $H_0$ measurements.\\


\textit{Methodology}---The standard siren technique uses the GW luminosity distance $D_{L,i}$ of an event $i$, combined with an independently determined source redshift, to infer $H_0$ via the flat-$\Lambda$CDM distance-redshift relation~\cite{Schutz1986,Chen2018}. For each of the 13 BBH-AGN associations identified in Ref.~\cite{Kumar2026a}, $H_0$ is obtained by inverting
\begin{equation}
  H_0 = \frac{c\,(1+z_{{\rm AGN},i})}{D_{L,i}}
        \left[\int_0^{z_{{\rm AGN},i}}\frac{dz'}{E(z')}\right]^{-1},
  \label{eq:H0}
\end{equation}
where $E(z)=[\Omega_m(1+z)^3+(1-\Omega_m)]^{1/2}$, $\Omega_m$ is the matter density, and $z_{{\rm AGN},i}$ is the AGN host redshift, corrected for the relativistic Doppler and gravitational shifts induced by the presence of the central SMBH, as derived in Ref.~\cite{Kumar2026a} following Refs.~\cite{Torres_Orjuela_2023,morton_rinaldi_2023}. Because each event is associated with a single, spectroscopically identified AGN host, $z_{{\rm AGN},i}$ enters as a fixed quantity rather than the marginalised host-redshift distribution $p(z)$ required by the galaxy-catalogue dark-siren method~\cite{Chen2018,Gray2020,Bom2024}; this removes the redshift marginalisation from the selection-function correction. The 13 per-event $H_0$ posteriors are combined, following the standard siren formalism~\cite{Chen2018,Gray2020,Mandel_2019}, as
\begin{equation}
  p(H_0\mid\boldsymbol{d}) =
  \frac{\displaystyle\prod_{i=1}^{N} p(H_0\mid d_i)}
       {\displaystyle\prod_{i=1}^{N} \beta_i(H_0)},
  \label{eq:combined}
\end{equation}
keeping the priors over the range $H_0\in[20,140]\,{\rm km\,s^{-1}\,Mpc^{-1}}$ and $\Omega_m\in[0.05,0.6]$. Here $\beta_i(H_0)$ is the selection-function correction accounting for GW detector sensitivity, analogous to the $\beta(H_0)$ term of \cite{Chen2018,Bom2024},
\begin{equation}
  \beta_i(H_0) = P_{\rm det}\!\left[D_L(z_{{\rm AGN},i},H_0)\right],
  \label{eq:beta}
\end{equation}
where $P_{\rm det}(D_L)$ is an injection-calibrated detection-probability curve constructed from the cumulative O1--O4b LVK search-sensitivity injection set~\cite{Essick:2025zed,Farr_2019}.

We report results from three runs differing only in the prior assumed on the source distance during GW parameter estimation. Run~1 adopts a deliberately wide prior that brackets both the Planck and SH0ES distance scales, so that the inference is not prejudiced toward either Cosmology; this is our primary, methodology-unbiased result. Run~2 exploits the precision of the AGN-derived redshift by tightening this prior around the value predicted at each event's $z_{{\rm AGN},i}$ under the Planck Cosmology, while Run~3 repeats this construction under the SH0ES Cosmology. Runs~2 and~3 quantify the precision recoverable under such an anchoring scheme. Runs~1--3 taken together therefore bound the trade-off between cosmological generality and statistical precision intrinsic to AGN-redshift standard sirens.\\

\textit{Results \& Discussion.}---
Table~\ref{tab:clean13_P6_full} lists the per-event $H_0$ posteriors for Run~1, together with the 13 BBH-AGN events combination and its sequential extension by other standard sirens. Individual events span $H_0=70.7$--$81.1\,{\rm km\,s^{-1}\,Mpc^{-1}}$ (median), with no individual event in tension with either Planck or SH0ES beyond $0.9\sigma$. For comparison, the Fisher information of individual 13 BBH-AGN events ranges from $0.88$--$1.51\times$ that of GW170817 alone (Table~\ref{tab:clean13_P6_full}), indicating that a majority of the AGN-associated events are individually as constraining as, or more constraining than, the single BNS (GW170817) detected to date.

\renewcommand{\arraystretch}{1.5}
\begin{table*}
\centering
\caption{$H_0$ constraints from the 13 individual BBH--AGN associations in Run~1, together with their joint combination and sequential extension by external standard sirens. For each row, $H_0$ is reported as the posterior median with 68.3\% and 90\% symmetric credible intervals, $^{+u68;\,+u90}_{-l68;\,-l90}$; $\delta_{H_0}=(\sigma_{68}/H_0)\times100$ is the fractional precision [\%]; $\sigma_{\rm Pl}$ and $\sigma_{\rm SH}$ are the quadrature tension with Planck\,2018 and SH0ES\,2024, respectively; and $\mathcal{R}_{\mathcal{F}}=\mathcal{F}/\mathcal{F}_{\rm BNS}$ is the Fisher information relative to the bright siren GW170817. The Bom et al.~\cite{Bom2024} row combines their 15 galaxy-catalogue dark sirens \emph{together with} GW170817 as a bright siren.}
\label{tab:clean13_P6_full}
\footnotesize
\setlength{\tabcolsep}{4pt}
\begin{tabular}{|l|c|c|c|c|c|}
\hline
GW Event & $H_0$ [km\,s$^{-1}$\,Mpc$^{-1}$] & $\delta_{H_0}$\,[\%] & $\sigma_{\rm Pl}$\,[$\sigma$] & $\sigma_{\rm SH}$\,[$\sigma$] & $\mathcal{R}_{\mathcal{F}}$ \\
 & $^{+u68;\,+u90}_{-l68;\,-l90}$ & & & & \\
\hline
GW240813\_034548 & $71.46^{+13.71;\,+25.97}_{-9.86;\,-15.15}$ & 16.49 & 0.34 & 0.13 & 1.51 \\
\hline
GW240807\_214559 & $76.03^{+15.87;\,+29.82}_{-11.54;\,-17.31}$ & 18.03 & 0.63 & 0.22 & 1.11 \\
\hline
GW190412\_053044 & $73.15^{+13.71;\,+25.73}_{-9.86;\,-14.91}$ & 16.11 & 0.49 & 0.01 & 1.51 \\
\hline
GW230922\_020344 & $71.46^{+14.43;\,+26.93}_{-10.34;\,-15.39}$ & 17.33 & 0.33 & 0.12 & 1.36 \\
\hline
GW231206\_233901 & $70.74^{+13.95;\,+26.69}_{-10.10;\,-14.91}$ & 17.00 & 0.28 & 0.19 & 1.45 \\
\hline
GW230630\_234532 & $75.31^{+16.11;\,+30.54}_{-11.06;\,-16.59}$ & 18.04 & 0.58 & 0.17 & 1.13 \\
\hline
GW190424\_180648 & $73.39^{+15.63;\,+29.82}_{-11.06;\,-16.59}$ & 18.19 & 0.45 & 0.03 & 1.17 \\
\hline
GW190514\_065416 & $73.63^{+15.39;\,+29.10}_{-11.06;\,-16.59}$ & 17.96 & 0.47 & 0.05 & 1.20 \\
\hline
GW190909\_114149 & $77.96^{+17.56;\,+32.95}_{-12.26;\,-18.28}$ & 19.13 & 0.71 & 0.33 & 0.94 \\
\hline
GW231123\_135430 & $76.51^{+16.59;\,+31.02}_{-11.54;\,-17.56}$ & 18.39 & 0.65 & 0.25 & 1.06 \\
\hline
GW190521\_030229 & $81.08^{+18.04;\,+33.19}_{-12.75;\,-19.48}$ & 18.98 & 0.89 & 0.52 & 0.88 \\
\hline
GW190803\_022701 & $72.91^{+15.39;\,+29.34}_{-10.82;\,-16.35}$ & 17.98 & 0.42 & 0.01 & 1.22 \\
\hline
GW200220\_124850 & $74.11^{+16.11;\,+30.30}_{-11.06;\,-16.59}$ & 18.33 & 0.49 & 0.08 & 1.13 \\
\hline
\hline
\rowcolor{green!15} 13 BBH-AGN  & $70.50^{+3.37;\,+5.77}_{-2.89;\,-4.81}$ & 4.43 & 0.98 & 0.76 & 21.4 \\
\hline
\rowcolor{green!15} 13 BBH-AGN $+$ BNS~\cite{Abbott2017Nature} & $70.31^{+3.00;\,+5.26}_{-2.85;\,-4.51}$ & 4.17 & 0.98 & 0.87 & 24.4 \\
\hline
\rowcolor{green!15} 13 BBH-AGN $+$ Bom~\cite{Bom2024} & $69.71^{+2.55;\,+4.21}_{-2.40;\,-3.90}$ & 3.55 & 0.91 & 1.23 & 34.1 \\
\hline
\end{tabular}
\end{table*}

\begin{figure}
\centering
\includegraphics[width=\columnwidth]{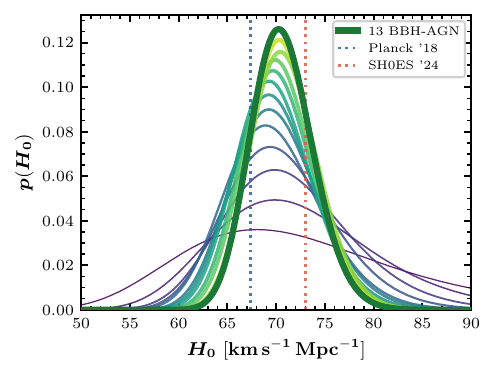}
\caption{Cumulative build-up of the joint $H_0$ posterior for Run~1 as the 13 BBH-AGN events are added in ascending $z_{{\rm AGN},i}$ order (dark-to-light gradient); the final 13-event posterior is highlighted (\textit{13 BBH-AGN}). Vertical dotted lines mark the Planck\,2018 and SH0ES\,2024 values.}
\label{fig:cumulative_buildup}
\end{figure}

Figure~\ref{fig:cumulative_buildup} shows the cumulative build-up of the $H_0$ posterior when combining the 13 independent BBH-AGN bright sirens, illustrating how the constraint tightens from any single event's $\sim$16-19\% fractional precision (cf. Table~\ref{tab:clean13_P6_full}) to the full-sample result, shown as the highlighted \textit{13 BBH-AGN} curve. Combining all 13 BBH-AGN associations yields
\begin{equation}
  H_0 = 70.50^{+3.37}_{-2.89}\,({\rm stat})\,\pm\,1.56\,({\rm cal})\;{\rm km\,s^{-1}\,Mpc^{-1}},
  \label{eq:h0_primary}
\end{equation}
a $4.4\%$ statistical fractional precision from the 13 BBH-AGN alone (Table~\ref{tab:clean13_P6_full}, 13 BBH-AGN row). Our result is driven entirely by the 13 BBH-AGN sample and does not rely on the BNS bright siren GW170817~\cite{Abbott2017Nature}; its subsequent inclusion below only marginally tightens the constraint. With the calibration systematic contributing an additional $2.2\%$ uncertainty. The corresponding measurement of the matter density yields $\Omega_m=0.307^{+0.263}_{-0.232}$ (90\% symmetric CI) for the combined 13 BBH-AGN bright sirens.

\begin{figure}
\centering
\includegraphics[width=\columnwidth]{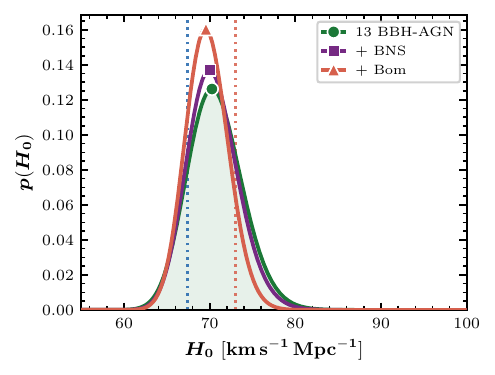}
\caption{The 13 BBH-AGN posterior of Fig.~\ref{fig:cumulative_buildup}, overlaid with its sequential combination with the bright siren GW170817 ($+$ BNS) and with the Bom et al.\ dark+bright (GW170817) siren sample ({$+$ Bom})~\cite{Bom2024}; markers denote each curve's MAP value. Vertical dotted lines mark Planck\,2018 and SH0ES\,2024.}
\label{fig:siren_overlay}
\end{figure}

Figure~\ref{fig:siren_overlay} shows the joint 13 BBH-AGN posterior alongside its sequential combination with the bright siren GW170817~\cite{Abbott2017Nature} and with the Bom et al.\ dark+bright (GW170817) siren sample~\cite{Bom2024} (Table~\ref{tab:clean13_P6_full}, bottom two rows). Each addition tightens the constraint monotonically: $H_0=70.31^{+3.00}_{-2.85}$ ($4.2\%$) with GW170817, and $H_0=69.71^{+2.55}_{-2.40}\,{\rm km\,s^{-1}\,Mpc^{-1}}$ ($3.5\%$) once the Bom et al.\ combination is folded in, with $\sigma_{\rm Planck}=0.91$ and $\sigma_{\rm SH0ES}=1.23$ -- a mild, sub-$1\sigma$ pull toward Planck\,2018 that does not constitute a meaningful preference for either Cosmology. The Fisher-information ratio $\mathcal{R}_{\mathcal{F}}$ climbs correspondingly from $21.4$ (13 BBH-AGN) to $34.1$ (13 BBH-AGN$+$Bom), confirming that each successive siren genuinely sharpens the constraint relative to GW170817 rather than simply diluting it.

The precision achieved in Run~1 represents a substantial improvement over both galaxy-catalogue dark-siren measurements and the current LVK population-based constraint, despite using an order of magnitude fewer events. Bom et al.~\cite{Bom2024} combine 15 galaxy-catalogue dark sirens to obtain $H_0=70.4^{+13.6}_{-11.7}\,{\rm km\,s^{-1}\,Mpc^{-1}}$ ($18\%$), tightening to $6\%$ only after folding in GW170817 with VLBI and peculiar-velocity corrections to the viewing angle. The LVK's own GWTC-5.0 analysis~\cite{theligoscientificcollaboration2026gwtc50constraintscosmicexpansion}, combining $235$ dark-siren events (galaxy catalogue plus spectral-siren mass-feature information) with GW170817, reaches $H_0=71.0^{+9.0}_{-7.1}\,{\rm km\,s^{-1}\,Mpc^{-1}}$ ($11\%$); the dark-siren contribution alone, before adding GW170817, is only $19\%$ precise. Our 13 BBH-AGN sample alone already matches or exceeds the precision of both of these fully combined results. Moreover, the results in Refs.~\cite{Bom2024,theligoscientificcollaboration2026gwtc50constraintscosmicexpansion} are dominated by GW170817, while our constraint, in contrast, is driven entirely by the AGN-associated BBH sample itself, with no reliance on any BNS bright siren. Our result combined with Bom's dark sirens and the BNS bright siren (GW170817) result approaches with $3.5\%$ a factor of two improvement over GWTC-5.0's full $236$-event combination. This gain follows directly from replacing galaxy-catalogue redshift marginalisation -- which must average over $\mathcal{O}(10^2$--$10^5)$ candidate host galaxies per event~\cite{theligoscientificcollaboration2026gwtc50constraintscosmicexpansion} -- with a single, spectroscopically confirmed AGN host per event: each 13 BBH-AGN association supplies as much redshift information as an entire galaxy catalogue's worth of marginalisation would for a comparable dark siren. Critically, this is achieved by our Cosmology-agnostic Run~1 result alone, with no assumption about the background Cosmology.

Runs~2 and~3 illustrate what this AGN-redshift precision can further deliver once paired with a correctly-anchored distance measurement, isolating how much of Run~1's uncertainty is set by its wide, Cosmology-bracketing construction rather than by any limitation of the AGN-siren method itself. Run~2 (Planck-anchored) yields $H_0=67.62^{+0.72}_{-0.72}\;{\rm km\,s^{-1}\,Mpc^{-1}}$ ($1.1\%$), and Run~3 (SH0ES-anchored) yields $H_0=72.91^{+0.72}_{-0.72}\;{\rm km\,s^{-1}\,Mpc^{-1}}$ ($1.0\%$), each recovering its respective anchor as expected by construction; the two report the same nominal $\pm0.72\,{\rm km\,s^{-1}\,Mpc^{-1}}$ precision but visibly different posterior shapes (68.3\% MAP: $67.62^{+0.48}_{-0.96}$ vs.\ $72.91^{+0.48}_{-1.20}$), reflecting the larger fractional distance uncertainty at the smaller SH0ES-implied luminosity distance. Sequentially adding GW170817 and the Bom et al.\ dark+bright (GW170817) combination only marginally affects the results of both runs (Run~2: $67.61^{+0.60}_{-0.60}$, $0.9\%$; Run~3: $72.72^{+0.75}_{-0.75}$, $1.0\%$). The corresponding tension values give the anchoring away directly: Run~2's final stage yields $\sigma_{\rm Planck}=0.27$ alongside $\sigma_{\rm SH0ES}=4.62$, while Run~3's yields $\sigma_{\rm SH0ES}=0.23$ alongside $\sigma_{\rm Planck}=5.89$ -- each run reproduces its own anchor to sub-$1\sigma$ while apparently rejecting the opposing Cosmology beyond $4$--$6\sigma$, a signature of the anchored construction rather than an independent cosmological discriminant. Runs~1--3 together bound the precision attainable from AGN-redshift standard sirens: once the background Cosmology is independently fixed, a comparable sample can reach sub-percent $H_0$ precision, underscoring the potential of BBH-AGN sirens as the confirmed-association sample grows.\\

\textit{Conclusion.}---
Single BBH-AGN associations have previously been used to constrain $H_0$, using the candidate AGN counterpart to GW190521~\cite{Mukherjee:2020kki,Gayathri:2020mra,morton_rinaldi_2023}. We have presented the first $H_0$ measurement using a catalogue of spectroscopically confirmed, one-to-one BBH--AGN associations. Using only 13 such associations, our Cosmology-agnostic result, $H_0=70.50^{+3.37}_{-2.89}\,({\rm stat})\pm1.56\,({\rm cal})\;{\rm km\,s^{-1}\,Mpc^{-1}}$ ($4.4\%$ statistical precision), already exceeds the precision of existing dark-siren-only measurements: Bom et al.~\cite{Bom2024} obtain $18\%$ from 15 galaxy-catalogue dark sirens alone, and LVK's GWTC-5.0 obtains $19\%$ from $235$ dark sirens alone~\cite{theligoscientificcollaboration2026gwtc50constraintscosmicexpansion}. Notably, this result is driven entirely by the 13 BBH-AGN sample and not by the BNS bright siren GW170817. This gain is a direct consequence of replacing per-event marginalisation over $\mathcal{O}(10^2$--$10^5)$ candidate host galaxies with a single confirmed spectroscopic redshift: each AGN association supplies, in one event, the redshift information an entire galaxy catalogue must average over for a comparable dark siren.

The 13 BBH-AGN sample shows no significant preference between Planck and SH0ES, with a residual sub-$1\sigma$ pull toward Planck that strengthens marginally once external sirens are included but does not by itself arbitrate the Hubble tension. Anchoring the luminosity distance prior to either Cosmology in turn demonstrates the precision attainable once a background Cosmology is assumed. We show explicitly that the resulting $\gtrsim4\sigma$ apparent rejections of the opposing anchor reflect the prior's own self-consistency rather than independent cosmological discrimination.

As the number of confidently identified BBH--AGN associations grows over upcoming LVK observing campaigns and deeper EM counterpart searches, the proposed method provides a direct and unbiased determination of \(H_0\) to sub-percent precision—requiring only 65, 258, and 1027 pairs to reach 2\%, 1\%, and 0.5\% accuracy, respectively.


\begin{acknowledgments}
\textit{Acknowledgment.---}We thank Andrew L. Miller, Bangalore S. Sathyaprakash, Bernard F. Schutz,
and Rachel Gray for helpful comments and questions.
This work was supported by the National Science Foundation of China
(No.\ W2533010). This work used data from the LIGO--Virgo--KAGRA
Collaboration provided by the Gravitational Wave Open Science Center
(\url{gwosc.org}). LIGO is funded by the NSF; Virgo by CNRS, INFN,
and Nikhef through EGO; KAGRA by MEXT and JSPS (Japan), NRF and MSIT
(Korea), and NSTC (Taiwan). D.K.\ would also like to acknowledge the
LIGO Laboratory computing resources supported by NSF grants:
PHY-0757058 and PHY-0823459, and the Gwave cluster maintained by the
Institute for Computational and Data Sciences at Penn State University,
supported by NSF grants: OAC-2346596, OAC-2201445, OAC-2103662,
OAC-2018299, and PHY-2110594. This material is based upon work supported by NSF’s LIGO Laboratory which is a major facility fully funded by the National Science Foundation.
\end{acknowledgments}

\bibliography{bibliography}
\end{document}